\documentclass[aps,twocolumn,showpacs,preprintnumbers,floatfix]{revtex4-1}
\usepackage{graphicx}
\usepackage{epsfig}
\usepackage{dcolumn}
\usepackage[colorlinks,linkcolor=red,anchorcolor=green,citecolor=blue,CJKbookmarks=True]{hyperref}
\usepackage{bbm}
\begin{document}
	
\title{Strangeonium-like hybrids on the lattice}
\author{\small Yunheng Ma,${}^{1,2}$ Ying Chen,${}^{1,2}$\thanks{cheny@ihep.ac.cn},
		Ming Gong,${}^{1,2}$, and Zhaofeng Liu${}^{1,2}$}
\affiliation{\small
$^1$~Institute of High Energy Physics, Chinese Academy of Sciences, Beijing 100049, P.R. China \\
$^2$~School of Physics, University of Chinese Academy of Sciences, Beijing 100049, P.R. China
                }

\begin{abstract}
The strangeonium-like $s\bar{s}g$ hybrids are investigated from lattice QCD in the quenched approximation. In the Coulomb gauge, spatially extended operators are constructed for $1^{--}$ and $(0,1,2)^{-+}$ states with the color octet $s\bar{s}$ component being separated from the chromomagnetic field strength by spatial distances $r$, whose matrix elements between the vacuum and the corresponding states are interpreted as Bethe-Salpeter (BS) wave functions. In each of the $(1,2)^{-+}$ channels, the masses and the BS wave functions are reliably derived. The $1^{-+}$ ground state mass is around 2.1-2.2 GeV, and that of $2^{-+}$ is around 2.3-2.4 GeV, while the masses 
of the first excited states are roughly 1.4 GeV higher. This mass splitting is much larger than the expectation of 
the phenomenological flux-tube model or constituent gluon model for hybrids, which is usually a few hundred MeV. The BS wave functions with respect to $r$ show clear radial nodal structures of non-relativistic two-body system, which  
imply that $r$ is a meaningful dynamical variable for these hybrids and motivate a color halo picture of hybrids that the color octet $s\bar{s}$ is surrounded by gluonic degrees of freedom. In the $1^{--}$ channel, the properties of the lowest two states comply with those of $\phi(1020)$ and $\phi(1680)$. We have not obtained convincing information 
relevant to $\phi(2170)$ yet, however, we argue that whether $\phi(2170)$ is a conventional $s\bar{s}$ meson or a 
$s\bar{s}g$ hybrid within the color halo scenario, the ratio of partial decay widths $\Gamma(\phi \eta)$ and $\Gamma (\phi \eta')$ observed by BESIII can be understood by the mechanism of hadronic transition of a strangeonium-like meson along with the $\eta-\eta'$ mixing. 

\end{abstract}


\pacs{12.38.Gc, 13.30.Ce, 14.40.Rt} \maketitle

\section{Introduction}

The naive quark model describes hadrons as the $q\bar{q}$ mesons and $qqq$ baryons. Since quarks and gluons are 
the fundamental degrees of freedom of QCD, if gluons can act as building blocks similar to quarks to build up hadrons,
then in the phenomenological meaning, there may exist glueballs which are purely made up of gluons, and hybrids which are composed of quarks and gluons. Glueballs, hybrids, and multiquark states (tetraquarks and pentaquarks, etc.) are usually called exotic hadrons in contrast to the conventional $q\bar{q}$ mesons and $qqq$ baryons. Exotic hadrons are
long-standing hot topics of the theoretical and experimental studies of particle physics, especially in the present era when quite a lot of $XYZ$ particles are discovered by various experiments, which come out with exotic properties 
in their production and decay processes and can be candidates for exotic hadrons~\cite{Brambilla:2019esw}. As of $q\bar{q}g$ hybrids made up of a quark-antiquark pair and a gluon, the $J^{PC}=1^{-+}$ states are most interesting since this quantum number is prohibited from the conventional $q\bar{q}$ mesons. There have been many theoretical studies on hybrids from phenomenological point of view and the lattice QCD approach. It is found that the lowest $1^{-+}$ hybrid usually has a mass of about 1 GeV higher than the ground state vector meson with the same $q\bar{q}$ component. For example, the mass of the $1^{-+}$ hybrid with light flavors is estimated to be about 1.9 GeV, the strangeonium-like and the charmonium-like counterparts are roughly 2.1-2.3 GeV~\cite{Lacock:1996ny,Dudek:2011bn} and 4.1-4.3 GeV~\cite{Liu:2012ze}, respectively. Experimentally, there are no reliable candidates for $1^{-+}$ hybrids established yet. The vector charmonium-like state $Y(4260)$ (or $\psi(4230)$ named by PDG 2018~\cite{Tanabashi:2018oca}), due to its very different properties from the conventional charmonia and the closeness of its mass to that of the $1^{-+}$ $c\bar{c}g$ hybrid, has a possible assignment of $1^{--}$ hybrid~\cite{Zhu:2005hp}. $\phi(2170)$\cite{Tanabashi:2018oca}, also known as $Y(2175)$, was first observed by the BABAR Collaboration in the initial-state-radiation process $e^+e^- \to \gamma_{\rm ISR}\phi f_0(980)$ in 2006\cite{Aubert:2006bu} and was confirmed later by BES and Belle~\cite{Ablikim:2007ab,Shen:2009zze}.
The similarity of its property to $Y(4260)$ also motivates a $s\bar{s}g$ hybrid interpretation of $\phi(2170)$. 

Phenomenologically, $q\bar{q}g$ hybrids are usually studied in the constituent gluon model~\cite{hybrid-horn-1978}, where the gluon acts as an effective degree of freedom similar to the constituent quarks in the quark model, or the flux tube model where the gluon is taken as a transverse vibration mode of the flux-tube which binds the $q\bar{q}$ pair~\cite{hybrid-isgur-1985}. As far as the hybrids with a heavy quark-anti-quark pair $Q\bar{Q}$ is concerned, the gluonic excitations along the flux-tube are fast objects, such that in the Oppenheimer approximation~\cite{Juge:1997nc, Juge:1999ie, Braaten:2014qka}, their distribution obeys the cylinder symmetry along the $Q\bar{Q}$-axis and their motion effects on the $Q\bar{Q}$ can be taken as a centrifugal barrier apart from the binding linear potential. Based on the hybrid potentials simulated from 
lattice QCD, one can solve the Schr\"{o}dinger equation of the $Q\bar{Q}$ system to give predictions of the spectrum of hybrids with properly tuned parameters. 

Lattice QCD is an {\it ab initio} non-perturbative approach for the study of strong interaction in the low energy scale, and is applied extensively to the investigation of hybrids~\cite{Lacock:1996vy,Bernard:1997ib,Liao:2002rj,Bernard:2003jd,Mei:2002ip,Dudek:2009qf, Yang:2012gz,Dudek:2008sz,Dudek:2009kk}. The masses of the hybrids can be derived from the correlation functions of hybrid-like operators $\bar{q}{\vec \Gamma} q \circ \vec{B}$, where $\bar{q}q$ is in the color octet, $\vec{B}$ is the chromomagnetic field strength, $\Gamma$ represents specific combinations of $\gamma$ matrices and the symbol $\circ$ means any possible summation of the spatial indices of $\vec{\Gamma}$ and $\vec{B}$. A recent lattice calculation~\cite{Liu:2012ze} shows that there exists a $\{1^{--},(0,1,2)^{-+}\}$ charmonium-like supermultiplet with nearly degenerate masses around 4.2-4.4 GeV, which overlaps strongly to the hybrid-like operators. This observation implies that these states may have similar internal dynamics, while the spin-spin coupling of the $\bar{q}q$ and $\vec{B}$ gives the different quantum numbers. In our previous work~\cite{Ma:2019hsm}, the internal structure of this supermultiplet was investigated by calculating their Bethe-Salpeter (BS) wave functions from lattice QCD in the quenched approximation, where the spatially extended interpolating field operators $\bar{q}{\vec \Gamma} q(\vec{x},t) \circ \vec{B}(\vec{x}+{\vec r},t)$ is introduced in the Coulomb gauge, whose matrix element between the vacuum and a state is defined as the BS wave function. It is found that the BS wave functions of the states in this multiplet are very similar and show interesting nodal structure, which implies that the distance between the $c\bar{c}$ and the $\vec{B}$ operator is a meaningful dynamical variable for hybrids. 

In this work, we extend the above study strategy to strangeonium-like hybrids and also focus on the $\{1^{--},(0,1,2)^{-+}\}$ states, such that we can check if the similar situation to the $c\bar{c}g$ hybrids can also happen for $s\bar{s}g$ states. On the other hand, since the quantum number $1^{--}$ and $0^{-+}$ are permitted by the $q\bar{q}$ mesons, in these channels, we will also use the spatially extended $s\bar{s}$ operators with the quark fields having spatial separations to extract the related BS wave functions, from which we can investigate the internal structure of these states. By the comparison of these two kinds of BS wave functions, we may obtain some information on the possible different formation pattern of hybrids from the conventional mesons. As for $\phi(2170)$, 
since it can be either $3^3 S_1$, $2^3 D_1$ $s\bar{s}$ or a candidate for the vector $s\bar{s}g$ hybrid, its properties will be discussed based on the results of this study. 

This work is organized as follows: Section~\ref{sec:numerical} gives a detailed description of our lattice setup and the numerical strategy including the construction of the spatially extended operators, the data analysis procedure and the results of the spectrum and BS wave functions. The discussion of our results and the comparison of our results with those phenomenological studies will be found in Section~\ref{sec:discussion}. Section~\ref{sec:summary} is a summary.

\section{Numerical details}\label{sec:numerical}
The pure gauge configurations are generated through the tadpole-improved gauge action~\cite{Morningstar:1999rf,Chen:2005mg} on anisotropic lattices with the aspect ratio being $\xi=a_s/a_t=5$, where $a_s$ and $a_t$ are the spatial and temporal lattice spacing, respectively. Two lattices $L^3\times T=16^3\times 160(\beta=2.4)$ and $24^3\times 192(\beta=2.8)$ with different lattice spacings are used to check the discretization artifacts. The parameters of the gauge ensembles are listed in
Table~\ref{tab:lattice}, where $a_s$ values are determined from $r_0^{-1}=410(20)$ MeV. 
For the strange valence quark, we use the tadpole-improved clover action whose parameters are tuned carefully by requiring the dispersion relations of vector
and pseudoscalar mesons to be reproduced ~\cite{Su:2004sc}. As will be addressed in the following sections, we will use spatially
extended operators to calculate the relevant correlation functions, therefore, the configurations are first 
fixed to the Coulomb gauge through the standard gauge fixing procedure in lattice QCD studies before the quark propagators are computed.

\begin{table}[t]
	\centering \caption{\label{tab:lattice}
		The input parameters for the calculation. Values of the coupling $\beta$, anisotropy
		$\xi$, the lattice spacing $a_s$, lattice size, and the number of measurements are
		listed.$a_s/r_0$
		is determined by the static potential, the first error of $a_s$ is the statistical error and the
		second one comes from the uncertainty of the scale parameter $r_0^{-1}=410(20)$ MeV.}
	\begin{ruledtabular}
		\begin{tabular}{cccccc}
			$\beta$ &  $\xi$  & $a_s$(fm) & $La_s$(fm) & $L^3\times T$ & $N_{conf}$ \\\hline
			2.4  & 5 & 0.222(2) & 3.55 &$16^3\times 160$ & 500 \\
			2.8  & 5 & 0.138(1) & 3.31 &$24^3\times 192$ & 200 \\
		\end{tabular}
	\end{ruledtabular}
\end{table}

\subsection{Interpolation field operators}
The major goal of this study is to investigate the inner structure of $s\bar{s}g$ hybrids of the quantum numbers of $J^{PC}=1^{--}, (0,1,2)^{-+}$. We introduce two types of spatially extended operators as sink operators. The first 
type includes the following $s\bar{s}g$ operators  
\begin{eqnarray} \label{ssg-sink}
O^k_{1^{--}}(r, t) &=& \sum_{{\vec x}, |{\vec r}|=r} {\bar s}^a({\vec x}, t)\gamma_5 s^b({\vec x}, t)B^{ab}_ k({\vec x}+{\vec r}, t) \nonumber\\
O_{0^{-+}}(r, t) &=& \sum_{{\vec x}, |{\vec r}|=r} {\bar s}^a({\vec x}, t)\gamma_i s^b({\vec x}, t)B^{ab}_ i({\vec x}+{\vec r}, t) \nonumber\\
O^k_{1^{-+}}(r, t) &=& \sum_{{\vec x}, |{\vec r}|=r} {\bar s}^a({\vec x}, t)\gamma_i s^b({\vec x}, t)B^{ab}_ j({\vec x}+{\vec r}, t)\varepsilon_{ijk} \nonumber\\
O^k_{2^{-+}}(r, t) &=& \sum_{{\vec x}, |{\vec r}|=r} {\bar s}^a({\vec x}, t)\gamma_i s^b({\vec x}, t)B^{ab}_ j({\vec x}+{\vec r}, t)|\varepsilon_{ijk}|,\nonumber\\
\end{eqnarray}
where $i,j,k=1,2,3$ are the spatial indices, $a, b=1,2,3$ are color indices, and  $B_i^{ab}=\frac{1}{2}\varepsilon_{ijk}F^{ab}_{jk}$ is chromomagnetic field strength as mentioned. The summation over ${\vec r}$ with same distance $r$ makes the operators have correct quantum numbers. Note that on a hypercubic spacetime latice, the spin $J=2$ corresponds to $T_2\oplus E$ where $T_2$ and $E$ are the irreducible representation of the lattice symmetry group $O$, so the $O_{2^{-+}}^k(r,t)$ are the three
components of $T_2$. The possible ground hybrids in these four quantum numbers form a supermultiplet as we expect. Obviously, the two constituent quarks are localized at same space-time point, the gluon component is placed at another space point. The BS wave function we try to extract reflects the dynamics for these two parts.

Since the quantum numbers $J^{PC}=0^{-+}, 1^{--}$ are conventional ones for $q\bar{q}$ mesons, we also introduce the 
second type of spatially extended $s\bar{s}$ operators by splitting the strange quark field $s$ and its conjugate $\bar{s}$ by a spatial separation $\vec{r}$, which are expressed explicitly as 
\begin{eqnarray} \label{ss-sink}
P_{0^{-+}}(r,t) &=& \sum_{{\vec x}, |{\vec r}|=r}  {\bar s}({\vec x}, t)\gamma_5 s({\vec x}+{\vec r}, t) , \nonumber\\
P^k_{1^{--}}(r,t) &=& \sum_{{\vec x}, |{\vec r}|=r} {\bar s}({\vec x}, t)\gamma_k s({\vec x}+{\vec r}, t) 
\end{eqnarray}
where the summation over $\vec{r}$ with $|\vec{r}|=r$ is again to guarantee the correct $J^{PC}$.

In practice, we calculate the wall-source correlation functions of these operators. For example, we use the following 
wall-source operators for $1^{-+}$ states,
\begin{equation} \label{ssg-source}
O^{W,k}(\tau) = \sum_{{\vec y},{\vec z}} {\bar s}^a({\vec y}, \tau)\gamma_i B^{ab}_ j({\vec z}, \tau) s^b({\vec z}, \tau)\varepsilon_{ijk},
\end{equation}
where $\tau$ labels the source time slice. The wall-source operators for other $J^{PC}$ states vary accordingly. The wall source operators for the $s\bar{s}$ operators are $P^{W}(\tau)=\sum\limits_{\vec{y},\vec{z}}\bar{s}(\vec{y},\tau)\Gamma s(\vec{z},\tau)$ with $\Gamma=\gamma_5, \gamma_i$ for $0^{-+}$ and $1^{--}$, respectively. At last, we calculate correlation functions (for simplicity, we set $\tau=0$ and omit the subscripts and superscripts referring to specific symmetry channels and different spatial components) as
\begin{equation} \label{ssg-corr}
C(r, t) = \langle O^k(r,t)O^{W,k\dagger}(0) \rangle 
\end{equation}
After the intermediate state insertion, the correlation function $C(r,t)$ can be parameterized as 

\begin{eqnarray} \label{ssg-para}
C(r,t) &=& \frac{1}{N_c}\sum_n \frac{1}{2m_nL^3}\langle 0| O(r,t) |n\rangle \langle n|O^W(0)|0\rangle \nonumber\\ 
&=& \frac{1}{N_c}\sum_n \frac{1}{2m_nL^3}\langle 0| O(r,0) |n\rangle \langle n|O^W(0)|0\rangle e^{-m_nt} \nonumber\\ 
&\equiv& \sum_n \Phi_n(r) e^{-m_nt}
\end{eqnarray}
where $N_c$ is the degenerate degree of $r=|\vec{r}|$, $m_n$ is the mass of the $n$-th state, and $\Phi_n(r)$ is defined as the corresponding Bethe-Salpeter wave function up to an irrelevant constant factor. Note that $m_n$ is independent of $r$, so that if we fit $C(r,t)$ 
with different $r$ simultaneously through Eq.(~\ref{ssg-para}), then we can obtain $m_n$ and $\Phi_n(r)$ altogether. To be specific, if $n_r$ different $C(r,t)$'s with different $r$ are considered and $N$ mass terms are involved in the fitting model, then the number of the parameters to be fitted is $N\cdot n_r+N$. Since we usually have 20-30 statistically meaningful data points for each $C(r,t)$, the number of the degree of freedom is large enough 
in the fitting procedure.   

\subsection{Results of $1^{-+}$ and $2^{-+}$ states}
We start with the $1^{-+}$ channel, since this $J^{PC}$ is a typical exotic quantum number which cannot be assigned to a 
$q\bar{q}$ meson in the quark model. After the correlation functions $C(r,t)$ are calculated on the $\beta=2.4$ and $\beta=2.8$ lattices, Eq.~\ref{ssg-para} is adopted for us to perform the data analysis where we use $N=3$ mass terms. On the $\beta=2.4$ lattice, the $r$ range is from 0 to $0.9$ fm (converted through the lattice spacing $a_s$ in Table~\ref{tab:lattice}) and the upper limit of the fit window $[t_{\rm min},t_{\rm max}]$ is uniformly set to be $t_{\rm max}=20$ for all the $C(r,t)$, while the $t_{\rm min}$ varies from 6 to 3. On the $\beta=2.8$ lattice, the $r$ range is up to $0.8$ fm and $t_{\rm max}$ is set to $t_{\rm max}=30$ with $t_{\rm min}$ varying from 9 to 6. 
On each lattice, we carry out a simultaneous correlated fit to all the $C(r,t)$'s with the jackknife covariance matrix. Table~\ref{tab:1mp-spec} shows the fit results of the masses $m_n$ with $n=1,2,3$ from different time window $[t_{\rm min},t_{\rm max}]$ as well as the $\chi^2$'s per degree of freedom ($\chi^2/{\rm dof}$) which are around one 
and indicate that the fits are reasonable. The masses of the lowest three states are stable to some extent at different $t_{\rm min}$ and are thereby reliable. For the lowest two states, the mass values on the $\beta=2.4$ lattice are mildly larger (roughly 100 MeV larger) than those on the $\beta=2.8$ lattice. This kind of difference might be attributed to the finite lattice spacing effects and that the strange quark mass parameters on the two lattices are not tuned to be exactly the same in the sense of the physical meaning. Combining the results from the two lattices, we can obtain that the mass of the lowest $1^{-+}$ $s\bar{s}g$ state is around 
2.2 GeV. On the other hand, the mass splitting of the lowest two states is around 1.4 GeV, which is almost the same as that of $1^{-+}$ charmonium-like hybrids, and thus shows to some extent the quark mass independence of this mass splitting.   

\begin{table}[t]
	\centering 
	\caption{\label{tab:1mp-spec}
	 The fitted masses $m_n$ of the $1^{-+}$ states with $n=1,2,3$ from different time window $[t_{\rm min},t_{\rm max}]$  as well as the $\chi^2$ per degree of freedom ($\chi^2/{\rm dof}$) on the  $\beta=2.4$ and $\beta=2.8$ lattices. All the masses are converted to the values in physical units through the lattice spacing $a_s$ in Table~\ref{tab:lattice}.}
 	\begin{ruledtabular}
	\begin{tabular}{ccccc}
		$t_{\rm min}$ & $\chi^2/{\rm dof}$ & $m_1$ (GeV) & $m_2$ (GeV) & $m_3$ (GeV) \\\hline
		        &$\beta=2.4$ &$t_{\rm max}=20$ &($s\bar{s}g$)&\\\hline
		6       &   1.12             & 2.232(22)	 &	 3.56(21)    & 7.5(2.7) \\
		5       &   1.23             & 2.228(22)	 &  3.61(26)    & 4.9(7)   \\
		4       &   1.36             & 2.248(13)    &  3.71(11)    & 5.7(5)   \\
		3       &   1.38             & 2.255(09)	 &  3.65(07)    & 5.4(2)   \\\hline
		        &$\beta=2.8$ &$t_{\rm max}=30$ &($s\bar{s}g$)&\\\hline
		9       &   1.53             & 2.099(16)	 &  3.55(07)    & 7.7(7)   \\
		8       &   1.41             & 2.168(15)	 &  3.78(13)    & 5.1(3)   \\	
		7       &   1.47             & 2.100(15)     &  3.40(07)    & 5.8(2)   \\	
		6       &   1.26             & 2.110(13)	 &  3.47(06)    & 5.5(1)   \\	
	\end{tabular}
	\end{ruledtabular}
\end{table}

\begin{figure}[t!]
	\includegraphics[height=5.5cm]{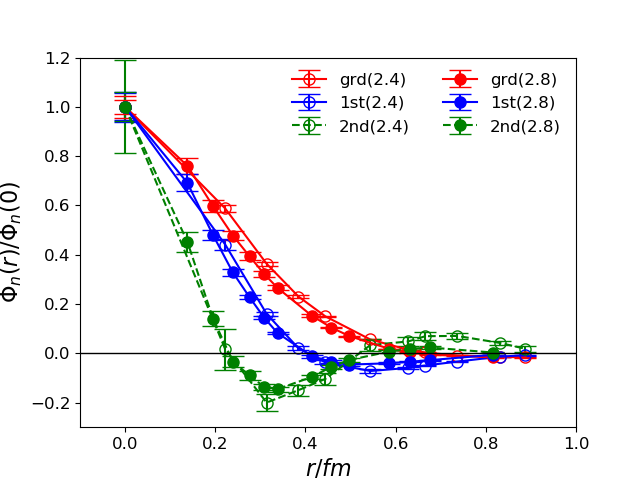}\quad
	\caption{\label{fig:wavefun1mp} The BS wave functions $\Phi_n(r)$ (normalized as $\Phi_n(0)=1$) of the lowest two $1^{-+}$ states. $r$ is the spatial separation between the $s\bar{s}$ component and the chromomagnetic operator $B_i$ and is converted to the value in physical units. Open and filled data points are the result of $\beta=2.4$ and $\beta=2.8$, respectively.}
\end{figure}

\begin{figure}[t!]
	\includegraphics[height=5.5cm]{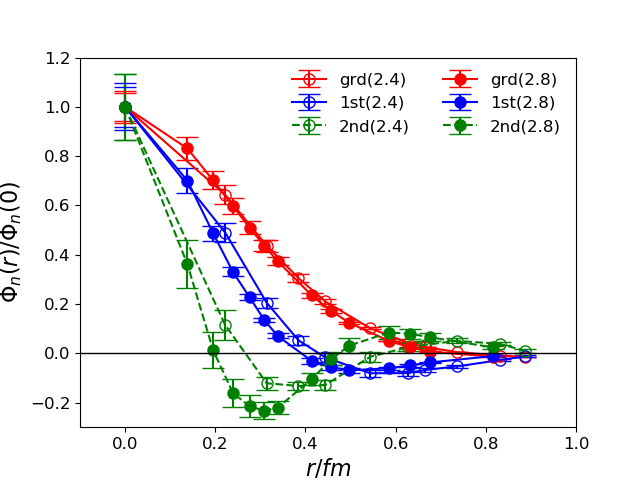}\quad
	\caption{\label{fig:wavefun2mp} The BS wave functions $\Phi_n(r)$ (normalized as $\Phi_n(0)=1$) of the lowest two $2^{-+}$ states. $r$ is the spatial separation between the $s\bar{s}$ component and the chromomagnetic operator $B_i$ and is converted to the value in physical units. Open and filled data points are the result of $\beta=2.4$ and $\beta=2.8$, respectively.}
\end{figure}

Along with the masses, the BS wave functions of these states $\Phi_n(r)$ can be extracted from the joint fit to $C(r,t)$'s, as shown in Fig.~\ref{fig:wavefun1mp}. The radial separation $r$ is converted to the value in the physical units and the wave functions on the two lattices are compatible with each other. The BS wave functions manifest clear nodal structure along the $r$ direction: the BS wave function $\Phi_1(r)$ of the ground state has no radial node, that of the first excited state ($\Phi_2(r)$) has one node, while that of the third state has two nodes. These nodal structures are very similar to the non-relativistic two-body Schr\"{o}dinger wave functions in a central potential. Note that $r$ is the spatial separation between the $s\bar{s}$ component and the color chromomagnetic field strength $\vec{B}$, the $r$ behaviors of the wave functions of the excitations may imply that 
within the $1^{-+}$ $s\bar{s}g$ hybrid, the relative movement between the $s\bar{s}$ and the gluonic degrees of freedom can be viewed qualitatively as a two-body system with $r$ being a physically meaningful dynamical variable.      
The same data analysis strategy is applied to the $2^{-+}$ channel. The fitted masses of the lowest three states are listed in Table~\ref{tab:2mp-spec}, and the wave functions are shown in Fig.~\ref{fig:wavefun2mp}. In comparison with the case of $1^{-+}$, the masses of $2^{-+}$ states are a little higher (100-200 MeV higher for the ground states) than their $1^{-+}$ counterparts but the pattern of the spectrum is similar. For the BS wave functions, the $r$ behavior of the $1^{-+}$ and $2^{-+}$ states are very alike. These observations support that the $1^{-+}$ and $2^{-+}$
states have almost the same inner structure and dynamics, while the small mass difference can be attributed to the 
different couplings between the spin of the $s\bar{s}$ subsystem and the gluonic degrees of freedom. This meets our expectation that $1^{-+}$ and $2^{-+}$ states with the nearly degenerate mass can be in the same supermultiplet. Of course, the possibility exists that these $2^{-+}$ states be the conventional $s\bar{s}$ mesons since $J^{PC}=2^{-+}$ is permitted for a $q\bar{q}$ system. However, the masses we obtain are much higher than those of ${}^1 D_2$ $s\bar{s}$ states in the quark model. On the other hand, a previous lattice study~\cite{Yang:2012mya} on charmonium states found that the $2^{-+}$ $c\bar{c}g$ operator couples almost exclusively to a state of mass 4.4 GeV instead of the expected $1^1D_2$ charmonium state $\eta_{c2}$ 
whose mass is around 3.8 GeV. If this is also the case for the $s\bar{s}$ states, the hybrid assignment is favorable for the $2^{-+}$ states we obtain in this work.   

\begin{table}[t]
	\centering \caption{\label{tab:2mp-spec}
		The fitted masses $m_n$ of the $2^{-+}$ states with $n=1,2,3$ from different time window $[t_{\rm min},t_{\rm max}]$  as well as the $\chi^2$ per degree of freedom ($\chi^2/{\rm dof}$) on the  $\beta=2.4$ and $\beta=2.8$ lattices. All the masses are converted to the values in physical units through the lattice spacing $a_s$ in Table~\ref{tab:lattice}.}
	\begin{ruledtabular}
		\begin{tabular}{ccccc}
		$t_{\rm min}$ & $\chi^2/{\rm dof}$ & $m_1$ (GeV) & $m_2$ (GeV) & $m_3$ (GeV) \\\hline
	    			  &$\beta=2.4$ &$t_{\rm max}=20$ &($s\bar{s}g$)&\\\hline
		      6       &   1.27             & 2.416(41)	 &	3.65(23)    & 7.2(2.4)   \\
		      5       &   1.37             & 2.406(41)	 &  3.60(25)    & 5.5(8)   \\
			  4       &   1.47             & 2.442(26)     &  3.69(19)    & 4.9(3)   \\
			  3       &   1.56             & 2.426(19)	 &  3.60(10)    & 5.0(1)   \\\hline
	                  &$\beta=2.8$ &$t_{\rm max}=30$ &($s\bar{s}g$)&\\\hline
			  9       &   1.47             & 2.361(23)	 &  3.89(09)    & 12.2(2.4)   \\
			  8       &   1.24             & 2.321(34)	 &  3.64(18)    & 5.5(5)   \\	
			  7       &   1.37             & 2.341(27)    &  3.71(12)    & 6.0(3)   \\	
			  6       &   1.40             & 2.359(19)	 &  4.00(08)    & 6.2(2)   \\	
		\end{tabular}
	\end{ruledtabular}
\end{table}

\subsection{Results of $0^{-+}$ and $1^{--}$ states}
The $0^{-+}$ and $1^{--}$ are conventional quantum numbers for $q\bar{q}$ mesons, and the mesons with these quantum numbers are usually assigned to be $n^1S_0$ and $n^3S_1$ states in the quark model. Therefore, we start with the analysis of the wall-source correlation functions $C(r,t)$ of $s\bar{s}$ operators with the $s$ and the $\bar{s}$ field separated by a spatial distance $r$.  We also use the function form of Eq.~(\ref{ssg-para}) with $N=3$ mass terms. The upper bound of the fit window is fixed to $t_{\rm max}=40$ and 30, and the lower bound goes gradually down to $t_{\rm min}=5$ and 7 for $\beta=2.4$ and $\beta=2.8$, respectively. The fitted masses of $1^{--}$ states are listed in Table~\ref{tab:1mm-spec} and those of $0^{-+}$ states are listed in Table~\ref{tab:0mp-spec}. As mentioned before, we use the physical mass of $\phi(1020)$ to set the mass parameters of the strange quark in the fermion action on the two lattices with smaller gauge ensembles. However, the fitted mass of the ground state deviates from the physical mass a little bit, which means that the strange quark masses are not tuned so precisely as enough. Therefore, one should keep in mind this mild deviation when looking at the data in the tables.

In the $1^{--}$ channel, the masses of the ground state and the first excited state are roughly 1 GeV and 1.7 GeV, which are compatible with those of $\phi(1020)$ and $\phi(1680)$. On the $\beta=2.4$ lattice, the fitted mass $m_3$ of the third state is also stable with respect to $t_{\rm min}$, and the value is around 2.1 GeV which close to expected mass of the $3^3S_1$ state from the quark model. The $m_3$ on the $\beta=2.8$ lattice is also in this mass range but fluctuates more strongly versus $t_{\rm min}$. 
  
\begin{figure}[t!]
	\includegraphics[height=5.5cm]{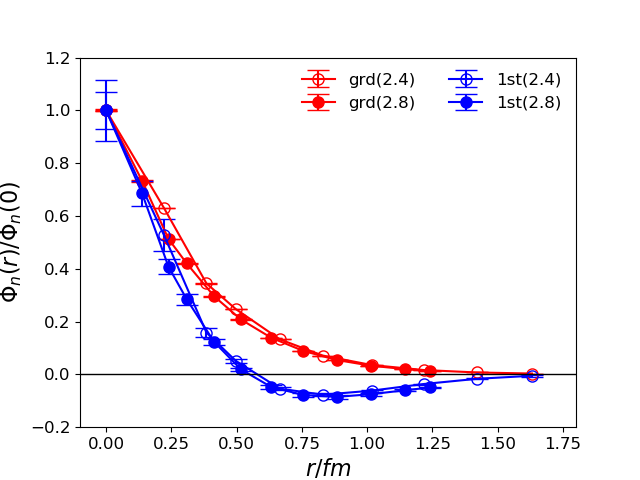}\quad
	\caption{\label{fig:wavefun0mpss} The BS wave functions $\Phi_n(r)$ (normalized as $\Phi_n(0)=1$) of the lowest two $0^{-+}$ states. $r$ is the spatial separation between the quark fields $s$ and $\bar{s}$ and is converted to the value in physical units. Open and filled data points are the result of $\beta=2.4$ and $\beta=2.8$, respectively.}
\end{figure}

\begin{figure}[t!]
	\includegraphics[height=5.5cm]{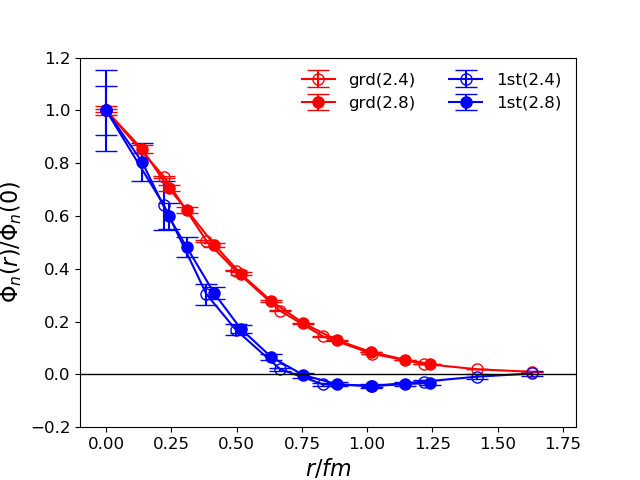}\quad
	\caption{\label{fig:wavefun1mmss}The BS wave functions $\Phi_n(r)$ (normalized as $\Phi_n(0)=1$) of the lowest two $1^{--}$ states. $r$ is the spatial separation between the quark fields $s$ and $\bar{s}$ and is converted to the value in physical units. Open and filled data points are the result of $\beta=2.4$ and $\beta=2.8$, respectively.}
\end{figure}

\begin{table}[t]
	\centering \caption{\label{tab:1mm-spec} The fitted masses $m_n$ of the $1^{--}$ states with two different types of operators and different time window $[t_{\rm min},t_{\rm max}]$ as well as the $\chi^2$ per degree of freedom ($\chi^2/{\rm dof}$) on the  $\beta=2.4$ and $\beta=2.8$ lattices. All the masses are converted to the values in physical units through the lattice spacing $a_s$ in Table~\ref{tab:lattice}.}
	\begin{ruledtabular}
		\begin{tabular}{ccccc}
		$t_{\rm min}$ & $\chi^2/{\rm dof}$ & $m_1$ (GeV) & $m_2$ (GeV) & $m_3$ (GeV) \\\hline
					  &$\beta=2.4$ &$t_{\rm max}=40$ &($s\bar{s}$)&\\\hline
			  8       &   0.84             & 1.013(1)	 &	1.753(77)   & 2.16(22) \\
			  7       &   0.81             & 1.013(1)	 &  1.787(93)  & 2.08(19) \\
			  6       &   0.83             & 1.014(1)  &  1.732(47)  & 2.13(12) \\
			  5       &   0.92             & 1.015(1)	 &  1.709(36)  & 2.11(08) \\
			\hline
					  &$\beta=2.4$ &$t_{\rm max}=45$ &($s\bar{s}g$)&\\\hline
			  16      &   1.40             & 1.011(1)  &  1.72(12)   &---\\
			  15      &   1.42             & 1.011(1)  &  1.66(10)   &---\\
			  14      &   1.42             & 1.009(1)  &  1.77(08)   &---\\
			  13      &   1.67             & 1.007(1)  &  1.83(06)   &---\\\hline
					  &$\beta=2.8$ &$t_{\rm max}=30$ &($s\bar{s}$)&\\\hline
			  10      &   1.93             & 1.001(5)	 &  1.634(114) & 2.17(29) \\
			  9       &   1.97             & 1.003(4)	 &  1.633(86)  & 2.02(18) \\	
			  8       &   2.11             & 0.999(3)  &  1.665(81)  & 2.30(20) \\	
			  7       &   2.39             & 0.998(3)	 &  1.668(52)  & 2.34(17) \\\hline
					  &$\beta=2.8$ &$t_{\rm max}=45$ &($s\bar{s}g$)&\\\hline
			  19      &   1.17             & 1.006(4)  & 1.51(6)     &---\\
			  18      &   1.39             & 1.003(3)  & 1.55(6)     &---\\
			  17      &   1.50             & 1.005(3)  & 1.51(5)     &---\\
			  16      &   1.45             & 0.998(2)  & 1.58(4)     &---\\
		\end{tabular}
	\end{ruledtabular}
\end{table}
 In the $0^{-+}$ channel, the ground state mass can be precisely determined with $m_1\approx 0.701$ GeV at $\beta=2.4$ and $0.651$ GeV at $\beta=2.8$. Since $s\bar{s}$ pseudoscalar meson (labeled as $\eta_s$) is not a physical state, we cannot compare our result to the physical value directly. There is a previous calculation from the $N_f=2+1$ full-QCD lattice formalism which gives the prediction $m_{\eta_s}=0.686(4)$ GeV~\cite{Davies:2009tsa}, which lies between our values from the two lattices. The deviation is mild and can be attributed to our less precise tuning of the strange quark mass parameter and the other systematic uncertainties. The mass of the first excited state is around 1.6-1.7 GeV, which is almost degenerate with that of the first excited $1^{--}$ state. There is no physical correspondence of this state
 yet, but one can compare it with the pseudoscalar $\eta(1295)/\eta(1475)$ but note that this state is a pure $s\bar{s}$ state which results in a higher mass. 
\begin{table}[t]
	\centering \caption{\label{tab:0mp-spec} The fitted masses $m_n$ of the $0^{-+}$ states with two different types of operators and different time window $[t_{\rm min},t_{\rm max}]$ as well as the $\chi^2$ per degree of freedom ($\chi^2/{\rm dof}$) on the  $\beta=2.4$ and $\beta=2.8$ lattices. All the masses are converted to the values in physical units through the lattice spacing $a_s$ in Table~\ref{tab:lattice}.}
	\begin{ruledtabular}
		\begin{tabular}{ccccc}
		$t_{\rm min}$ & $\chi^2/{\rm dof}$ & $m_1$ (GeV) & $m_2$ (GeV) & $m_3$ (GeV) \\\hline
					  &$\beta=2.4$ &$t_{\rm max}=40$ &($s\bar{s}$)&\\\hline
			 8        &   0.70             & 0.7012(2)	 &	1.690(34)   & 2.21(17) \\
			 7        &   0.69             & 0.7010(2)	 &  1.698(39)  & 2.12(14) \\
			 6        &   0.73             & 0.7010(2)   &  1.699(30)  & 2.12(10) \\
			 5        &   0.89             & 0.7014(2)	 &  1.669(22)  & 2.10(07) \\\hline
					  &$\beta=2.4$ &$t_{\rm max}=45$ &($s\bar{s}g$)&\\\hline
			 16       &   1.30             & 0.7008(3)    &  1.711(73)  &---\\
			 15       &   1.40             & 0.7007(3)    &  1.680(56)  &---\\
			 14       &   1.42             & 0.7007(3)    &  1.672(46)  &---\\
			 13       &   1.56             & 0.7009(3)    &  1.659(36)  &---\\\hline
					  &$\beta=2.8$ &$t_{\rm max}=30$ &($s\bar{s}$)&\\\hline   
			 10       &   2.59             & 0.6483(8)	 &  1.736(62)  & 2.84(34) \\
			 9        &   2.51             & 0.6505(8)	 &  1.703(76)  & 2.33(25) \\	
			 8        &   2.37             & 0.6512(8)   &  1.679(66)  & 2.23(20) \\	
			 7        &   2.35             & 0.6516(7)	 &  1.620(30)  & 2.46(11) \\\hline
					  &$\beta=2.8$ &$t_{\rm max}=45$ &($s\bar{s}g$)&\\\hline
			 19       &   1.09             & 0.6508(10)   &  1.621(85)  &---\\
			 18       &   1.03             & 0.6510(10)   &  1.557(67)  &---\\
			 17       &   1.04             & 0.6490(06)   &  1.711(50)  &---\\
			 16       &   1.06             & 0.6491(05)   &  1.740(39)  &---\\
		\end{tabular}
	\end{ruledtabular}
\end{table}
As mentioned above, the BS wave functions $\Phi_n(r)$ of $0^{-+}$ and $1^{--}$ states can be derived simultaneously with the masses, which are shown in Fig.~\ref{fig:wavefun0mpss} and ~\ref{fig:wavefun1mmss}. Note here $r$ stands for the separation between $s$ and $\bar{s}$ field. For the ground state and the first excited state in each channel, the $\Phi_n(r)$'s exhibit the expectation of the quark model that the wave function of the ground state has no radial node while that of the first excited state has one node. Therefore, the two states can be assigned to be the $1S$ and $2S$ state of a non-relativistic $s\bar{s}$ system. The behavior of the wave function of the third state is strange that it has no radial nodes even though it has two inflection points. Since we only use three mass terms to fit the correlation functions, the third state may has substantial contamination from higher states which may result in this phenomenon. So we do not take the results of the third state too seriously.    

\begin{figure}[t!]
	\includegraphics[height=5.5cm]{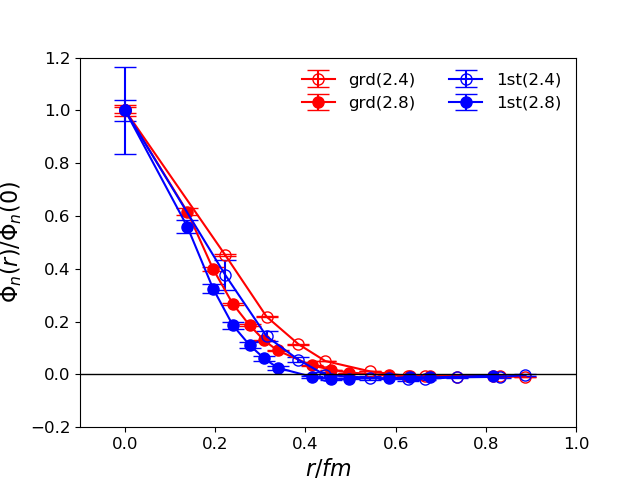}\quad
	\caption{\label{fig:wavefun1mmssg}The BS wave functions $\Phi_n(r)$ (normalized as $\Phi_n(0)=1$) of the lowest two $1^{--}$ states. $r$ is the spatial separation between the $s\bar{s}$ component and the chromomagnetic operator $B_i$ and is converted to the value in physical units. Open and filled data points are the result of $\beta=2.4$ and $\beta=2.8$, respectively. Since the two states can be assigned to be the $1S$ and $2S$ $s\bar{s}$ mesons, the similar $r$-behavior of their BS wave functions imply that this $r$ is not a typical dynamical variable for $s\bar{s}$ states.}
\end{figure}

\begin{figure}[t!]
	\includegraphics[height=5.5cm]{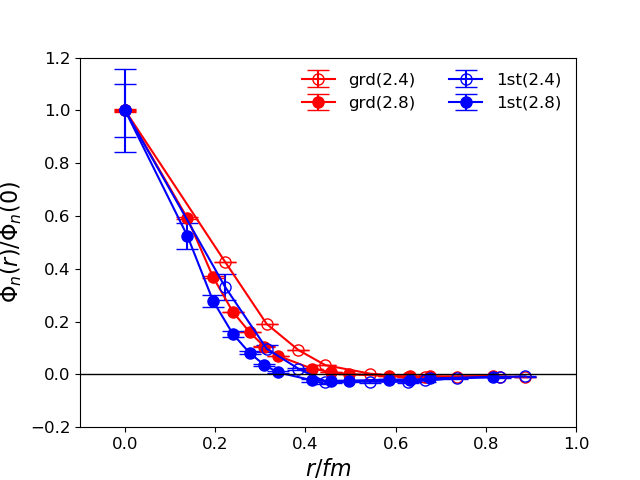}\quad
	\caption{\label{fig:wavefun0mpssg}Similar to Fig.~\ref{fig:wavefun1mmssg}, but for $0^{-+}$ states.}
\end{figure}

We also use the $s\bar{s}g$-type operators (in Eq.~(\ref{ssg-sink})) to explore the properties of the $0^{-+}$ and $1^{--}$ states. We use Eq.~(\ref{ssg-para}) of $N=2$ mass terms to fit the correlation functions in large time ranges ($t_{\rm max}=45$ for both lattices). The masses are shown in  Table~\ref{tab:1mm-spec} and Table~\ref{tab:0mp-spec} where one can see that they are consistent with those from the $s\bar{s}$ operators. Figure~\ref{fig:wavefun1mmssg} and ~\ref{fig:wavefun0mpssg} show the BS wave functions with $r$ being the spatial separation between the $s\bar{s}$ components and the chromomagnetic operator. The results of the two lattices are compatible with each other. It is interesting to see that, in $0^{-+}$ and $1^{--}$ channels, this kind of wave functions of the ground state ($1S$ state) and the first excited state ($2S$ state) lie almost upon each other and there are no sharp difference. In these two channels, given the masses of which are compatible with masses of the states obtained by $s\bar{s}$ operators, the ground states and the first excited states can be assigned 
to the $1S$ and $2S$ $s\bar{s}$ mesons. As such the similarity of the wave functions of $1S$ and $2S$ states with respect to the distance $r$ between the gluonic component and the $s\bar{s}$ component can be interpreted as follows: 
the $s$ (or $\bar{s}$) field along with the chromomagnetic field can be viewed as a dressed $s'$ (or $\bar{s}'$) field
in the fundamental representation of the color $SU(3)$ group, which annihilates the $s$ (anti)quark of the $s\bar{s}$ meson. In this sense, the $r$ reflects the spatial size of the dressed quark field and the $r$ fall-off 
does not have a dynamical significance. This is in contrast to the wave functions of $1^{-+}$ and $2^{-+}$ states
whose nodal structures imply that $r$ is a dynamical variable for hybrids.  

\section{Discussion}\label{sec:discussion}
We discuss the above results in this section. The BS wave functions of $1^{-+}$ and $2^{-+}$ states show the typical 
behaviors of the non-relativistic two-body Schr\"{o}dinger wave functions with a central potential, say, the correspondence of the spectrum and the nodal structure of the wave functions. We emphasize that 
the variable $r$ is the spatial distance between the $s\bar{s}$ component and the chromomagnetic field strength of the operators. Since $1^{-+}$ is an exotic quantum number for $q\bar{q}$ mesons, the states with this quantum number must be a hybrid meson with additional gluonic degrees of freedom. The similarity of the spectrum and the wave functions of $2^{-+}$ states to $1^{-+}$ ones signals that they are also hybrid states. In this sense, the wave functions imply that the $r$ can be a meaningful dynamical variable for $s\bar{s}g$ hybrid mesons. In a previous 
lattice study on $c\bar{c}g$ hybrids~\cite{Ma:2019hsm}, the same behaviors of the wave functions and the spectrum pattern have been observed for the $(0,1,2)^{-+}$ and $1^{--}$ supermultiplets, based on which a "color halo" picture has been proposed that a hybrid meson can be 
viewed as a relatively compact color octet $q\bar{q}$ pair surrounded by color octet gluonic degrees of freedom such that the wave functions depict the relative motion between the $q\bar{q}$ pair and the gluonic excitation. In a non-relativistic picture, the binding mechanism can be the potential between two effective color octet charges. Previous lattice studies~\cite{Bali:2000gf} show that the potential of two static color charge has a feature of Casimir scaling 
\begin{equation}
V_D(r)=V_{D,0}-C_D\frac{\alpha}{r} +\sigma_D r
\end{equation} 
where $D$ labels the color $SU(3)$ representation of the charge with $C_D$ being the eigenvalue of the second order Casimir operator of the color $SU(3)$, $V_{D,0}$ is the pontential constant, $\alpha$ is the coefficient of the Coulomb part, and $\sigma_D$ is the string tension which is related to the conventional 
string tension $\sigma$ between a static quark and antiquark pair by $\sigma_D =\frac{3}{4} C_D \sigma$ 
. For the octet charge in this work, this relation is $\sigma_D =9/4 \sigma$, which means the interaction between color octet objects is stronger than that of color triplet ones. This explains the observation that mass splitting (about 1.2 GeV) between the ground state and the first excited hybrid state is larger than that of $1S-2S$ mass splitting of $q\bar{q}$ states (around 0.6 GeV). 

The 'color halo' picture is conceptually different from the flux-tube picture of hybrids in the market~\cite{Juge:1997nc, Juge:1999ie, Braaten:2014qka, Ding:2006ya, Akbar:2020vuo}, where the 
quark and anti-quark is bound by an effective potential induced by the excitation of gluonic degrees of freedom. In the leading Born-Oppenheimer approximation,  
the $Q\bar{Q}$ of a heavy quarkonium-like hybrid can be viewed as static color sources, the excited gluonic degrees of freedom distribute along the $Q\bar{Q}$ axis and obey the cylinder symmetry, whose effect can be treated as an excited static potential denoted  by $\Lambda_\eta^\epsilon$, where $\Lambda=0,1, 2,\ldots$ is the projected total angular momentum of gluons with respect to the $Q\bar{Q}$ axis and is labeled as $\Sigma, \Pi, \Delta$ for $\Lambda=0,1,2$ and so on, $\eta $ represents the combined parity ($P$) and the charge conjugate ($C$)
of gluon excitations with $\eta=g,u$ for $P\otimes C=\pm$, respectively, and $\epsilon$ is the $P$ parity of the glue state. Therefore, the 
quantum number of a $Q\bar{Q}$ state with this kind of potential is 
\begin{equation}
P=\epsilon (-1)^{L+\Lambda+1}, C=\eta\epsilon (-1)^{L+S+\Lambda}
\end{equation}
where $\hat{L} =\hat{L}_{Q\bar{Q}}+\hat{J}_g$ with $\hat{L}_{Q\bar{Q}}$ being the orbital angular momentum of $Q\bar{Q}$ with respect to the midpoint of the $Q\bar{Q}$ axis, and $\hat{J}_g$ being the total angular momentum of gluons. The ground $\Sigma_g^+$ potential has $\Lambda=0, \epsilon=+$ and $\eta=+$ is actually the conventional 
static potential of $Q\bar{Q}$ of the Cornell type, and the $P$ and $C$ quantum nubmer reproduce the conventional quantum number. The $1^{--}$ and $(0,1,2)^{-+}$ hybrid supermultiplet is associated with the $\Pi^+_u (L=1)$ potential, such that the radial Shr\"{o}dinger equation is 
\begin{equation}
\frac{d^2}{dr^2} u(r)+2\mu[E-V_{\rm eff}(r)]u(r)=0
\end{equation}
where $r$ is the distance between $Q$ and $\bar{Q}$, $\mu$ is the reduced mass of the $Q\bar{Q}$ pair, and $u(r)$ is related to the radial wave function $\phi(r)$ by $u(r)=r\phi(r)$. The effective potential $V_{\rm eff}$ is 
\begin{equation}
V_{\rm eff}=V_{Q\bar{Q}}(r)+\frac{\langle \hat{L}_{Q\bar{Q}}^2\rangle}{2\mu r^2}
\end{equation}
with $\langle \hat{L}_{Q\bar{Q}}^2\rangle=L(L+1)-2\Lambda^2+\langle \hat{J}_g^2\rangle$ and $\langle \hat{J}_g^2\rangle=2$. Obviously the eigenvalues of $E$ is independent of the total spin $S$ of the $Q\bar{Q}$ pair. One can use the lattice results to determine $V_{Q\bar{Q}}$ and then solve the above equation to get the 
masses of the hybrids. We would not like to go into much details of the studies in this direction, but only mention that $\phi_n(r)$ behaves as a $P$-wave wave function in a central potential, and the mass splitting of the ground 
state and the first radial excited state is only a few hundred MeV for $1^{--}$ and $(0,1,2)^{-+}$ hybrid states~\cite{Braaten:2014qka}. 
Even though the above deduction is based on the heavy quarkonium-like hybrids, this picture has been also applied to the phenomenological studies of strangeonium hybrids~\cite{Ding:2006ya}. 

In contrast to the flux-tube picture, we observe that for $(1,2)^{-+}$ strangeonium hybrids, the mass splitting 
of the ground and the first excited states is around 1.2-1.4 GeV, which is much larger than the prediction of the flux-tube model, and the nodal structure shows 
up with respect to the spatial distance between the $s\bar{s}$ and the chromomagnetic field strength. Similar phenomenon also appears for charmonium-like hybrids without a clear quark mass dependence. It should be emphasized that even though the interpretation of the wave functions can be debatable, the pattern of the spectrum should be 
solid and model independent since it is derived directly from the lattice QCD calculation. 

As far as the $0^{-+}$ and $1^{--}$ channels are concerned, we get consistent results of the masses of the ground states
and the first excited state by using the $s\bar{s}$ type and $s\bar{s}g$ type operators. Since we use the physical 
mass of the $\phi(1020)$ meson to set the strange quark mass parameters, it is natural to almost reproduce the physical value of the mass of the vector ground state. The ground state mass of the pseudoscalar is around $650-700$ MeV, which is compatible with the previous lattice result of $\eta_s$. The masses of the first excited states 
in both channels are closely degenerate at $1.7$ GeV, and the mass of the first excited $s\bar{s}$ vector meson is in 
agreement with that of the $\phi(1680)$. On the other hand, the BS wave functions in both channels, defined through the dependence of  spatial distance between the $s$ and $\bar{s}$ quark field, show the expected radially nodal 
behavior of the non-relativistic $s\bar{s}$ two-body system. Therefore, the ground and the first excited states 
can be assigned to be the $1S$ and $2S$ $s\bar{s}$ mesons, respectively. We also obtain some information of the third state through the $s\bar{s}$ type operator in each channel, whose mass is around 2.1-2.3 GeV. For the vector channel, 
this mass value is close to the mass of $\phi(2170)$. However, since we only use three mass terms to do the data fitting, the third state may have substantial contaminations from higher states, the result is not that solid. When we use the $s\bar{s}g$ operator to study these two channels, we can only obtain the information of the lowest two states. At present, we have no decisive conclusion if there is a $1^{--}$ and $(0,1,2)^{-+}$ supermultiplet of the 
strangeonium hybrids. 

At last, we make some arguments on the $\phi(2170)$. Its mass is in the mass range of $3^3S_1$ and $2^3D_1$ $s\bar{s}$ predicted by the quark model. If there does exist a $1^{--}$ and $(0,1,2)^{-+}$ $s\bar{s}g$ hybrid multiplet with nearly degenerate masses around $2.1-2.3$ GeV, then $\phi(2170)$ can be also a candidate for the $1^{--}$ member. However, the assignment of its nature is still an open question. Till now, $\phi(2170)$ has been observed in many final states including $\phi(1020)$, such as the $\phi f_0(980)$, $\phi\pi\pi$, $\phi \eta$ and $\phi \eta'$ etc. In the $K^+K^-\pi\pi$ and $K^+K^-K^+K^-$ final states~\cite{Lees:2011zi, Ablikim:2019tpp}, there are also sizable components including $\phi(1020)$. If $\phi(2170)$ is a candidate for the $1^{--}$ $s\bar{s}g$ hybrid, this decay pattern can be understood within the 'color halo' picture of the hybrids: the binding between the color octet $s\bar{s}$ and the gluonic degrees of freedom can easily break up such that the $s\bar{s}$ component is neutralized to $\phi(1020)$ and 
the gluons are hadronized to light hadrons which are in the flavor singlet. Furthermore, in contrast to the hadronic
transition of conventional excited strangeonium states, these decays are less OZI suppressed due to the existing 
gluons within strangeonium hybrids.  Recently, the BESIII Collaboration reported the observation of $\phi(2170)$ in the process $e^+e^-\to \eta'\phi$ with the resonance parameters $M_R=2177.5\pm 4.8({\rm stat})\pm 19.5({\rm syst})$ MeV and $\Gamma_R=149.0\pm 15.6({\rm stat})\pm 8.9({\rm syst})$ MeV, and ${\rm Br}(\phi(2170)
\to \eta'\phi)\Gamma_{e^+e^-}$ is measured to be $7.1\pm 0.7({\rm stat})\pm 0.7({\rm syst})$ eV~\cite{Ablikim:2020coo}. Combining the result of ${\rm Br}(\phi(2170)
\to \eta\phi)\Gamma_{e^+e^-}=1.7\pm 0.7({\rm stat})\pm 1.3({\rm syst})$ eV, one has 
\begin{equation}
\frac{{\rm Br}(\phi(2170)
\to \eta\phi)\Gamma_{e^+e^-}}{{\rm Br}(\phi(2170)
\to \eta'\phi)\Gamma_{e^+e^-}}=0.23\pm 0.10({\rm stat})\pm 0.18({\rm syst}).
\end{equation}
This ratio is much larger than the predictions of the phenomenological studies based on the flux tube model or 
the constituent gluon model of hybrids with the mechanism that the flux tube or the constituent gluon breaks up into 
a light $q\bar{q}$ pair which reorganizes into two mesons with the original constituent $s\bar{s}$. However, this ratio can be explained directly from the flavor octet-singlet mixing and the kinetics. If $\phi(2170)$ is a $s\bar{s}g$ hybrid in the 'color halo' picture, then the decay $\phi(2170)\to \phi\eta(\eta')$ can take place as follows: a gluon emitted by the constituent strange quark (or antiquark) and the original gluon(s) couple to the flavor singlet component of the $\eta(\eta')$ meson. If $\phi(2170)$ is a higher excited $s\bar{s}$ meson, then 
the $\eta(\eta')$ is generated by two gluons emitted by the $s\bar{s}$ pair. Note that this process can be enhanced by the QCD axial anomaly. Since the decay dynamics is expected to be the same for the $\phi\eta$ and $\phi\eta'$ decay mode, the ratio of the partial widths can be attributed to the $\eta-\eta'$ mixing and the kinetic factor
\begin{equation}
\frac{\Gamma(\phi(2170)\to\phi\eta)}{\Gamma(\phi(2170)\to\phi\eta')}=\tan^2\theta \left(\frac{k_\eta}{k_{\eta'}}\right)^3
\end{equation}  
where $\theta$ is the flavor octet-singlet mixing angle of $\eta-\eta'$ system and $k_{\eta^{(')}}$ is the magnitude of the decay momentum. If we take the physical masses $m_\eta=547$ MeV and $m_{\eta'}=958$ MeV and the mixing angle $\theta$ varying between $-10^\circ$ and  $-20^\circ$, this ratio is estimated to be between $0.14$ and $0.58$ and compatible with the experimental value (the mixing 
angle $\theta$ is derived to be around $|\theta|\approx 13^\circ$ if using the central value 0.23). In other words, 
for $\phi(2170)$, this ratio may not be a good criterion to distinguish a hybrid assignment from a conventional 
$s\bar{s}$ meson.  
 	
\section{Summary}\label{sec:summary}
The strangeonium-like hybrids are investigated from lattice QCD in the quenched approximation. Two anisotropic lattices with different lattice spacings are used to check the finite $a_s$ effects. We construct spatially extended 
$s\bar{s}g$ operators with the $s\bar{s}$ component separated from the chromomagnetic field strength operator by a 
spatial distance $r$. We focus on the $1^{--}$ and $(0,1,2)^{-+}$ channels and calculate the corresponding  correlation functions based on these operators in the Coulomb gauge. The ground state mass of the $1^{-+}$ states is determined to be 2.1-2.2 GeV and that of the $2^{-+}$ states is about 200 MeV higher. These results are consistent with the previous lattice calculations and phenomenological studies. The masses of the first excited state are around 3.6 GeV in these two channels, such that the mass splittings of the first excited states and the ground states are roughly 1.2-1.4 
GeV, which is much higher than the predictions of the flux-tube model which is only a few hundred MeV.  The BS wave functions of these states, defined by the matrix elements of the operators mentioned above between the vacuum and the states, are extracted and show clear nodal structures in the $r$ direction, which manifest that $r$ is a meaningful 
dynamical variable reflecting the relative motion of the center-of-mass of the $s\bar{s}$ against the gluonic degrees of freedom. Both the spectrum and the wave functions of these $s\bar{s}g$ states have similar feature to their
$c\bar{c}g$ counterparts and comply with the 'color halo' picture of hybrids that the color octet $q\bar{q}$ pair is surrounded by gluons. 

In the $0^{-+}$ and $1^{--}$ channels, we use both the spatially extended $s\bar{s}$ and $s\bar{s}g$ operators to carry out the calculations. The ground state mass of the vector $s\bar{s}$ meson almost reproduce the mass of $\phi(1020)$ (note that we use the mass of the $\phi(1020)$ to set the strange quark mass parameters), and the ground state mass of the pseudoscalar is 650-700 MeV, which is in agreement with the $\eta_s$ mass determined by previous lattice calculations. In both channels, the masses of the first excited states are almost degenerate at around 1.7 GeV and compatible with the mass of $\phi(1680)$. The BS wave functions of these states with respect to the distance between $s$ and $\bar{s}$ are qualitatively similar to the nonrelativistic wave function of a two-body system in that  the BS wave function of the first excited state has a radial node. Therefore, the first excited state can be a $2S$ $s\bar{s}$ meson. In contrast, in each channel, the BS wave functions with respect to the spatial distance of the octet $s\bar{s}$ and the gluonic degrees of freedom have similar profile for the ground and the first excited state, which means this distance is less significant for $s\bar{s}$ mesons. 

We have not gotten solid results for the $3S$ $s\bar{s}$ mesons and the possible vector $s\bar{s}g$ hybrids, therefore we cannot give a convincing explanation of the $\phi(2170)$. Since the mass of $\phi(2170)$ is compatible with the quark model prediction of $3S$ $s\bar{s}$ meson and the predicted mass of the lowest $s\bar{s}g$ hybrids, both assignments of $\phi(2170)$ are possible. We argue that
if $\phi(2170)$ is either the $3S$ $s\bar{s}$ meson or a vector $s\bar{s}g$ hybrid within the 'color halo' picture discussed above, the ratio $\Gamma(\phi\eta)/\Gamma(\phi\eta')= 0.23\pm 0.10({\rm stat})\pm 0.18({\rm syst})$ can be understood by the hadronic transition of a strangeoium-like meson along with the $\eta-\eta'$ mixing. Anyway, the nature of $\phi(2170)$ is still an open question to be investigated by further experimental and theoretical studies.

\section*{ACKNOWLEDGEMENTS}
    This work is supported by the National Key Research and Development Program of China
	(No.2017YFB0203202) and the Strategic Priority Research Program of Chinese Academy of Sciences (No.XDC01040100 and XDB34030302). The numerical calculations are carried out on Tianhe-1A at the National
	Supercomputer Center (NSCC) in Tianjin and the GPU cluster at IHEP. We also
	acknowledge the support of the National Science Foundation of China (NSFC) under Grants
	No.11935017, No. 11775229, No. 11575196, No. 11575197, and No. 11621131001 (CRC 110 by DFG and NSFC). Y.C. is also supported by the CAS Center for Excellence in Particle Physics (CCEPP).

\bibliography{ref}

\end{document}